# Keyboards for inputting Japanese language -A study based on US patents

**Umakant Mishra**
Bangalore, India
umakant@trizsite.tk
http://umakant.trizsite.tk

**Contents**

## 1. Introduction

The most commonly used Japanese alphabets are Kanji, Hiragana and Katakana. The Kanji alphabet includes pictographs or ideographic characters that were adopted from the Chinese alphabet. Hiragana and Katakana are phonetic alphabets that do not include any characters common to each other or to Kanji. Hiragana is used to spell words of Japanese origin, while Katakana is used to spell words of western or other foreign origin. The term kana is used to refer to either hiragana or katakana.

## 1.1 The features of Japanese characters

The Japanese syllabary includes 108 syllables. There are some additional rarely used syllables, such as the "small" versions of the vowel syllables that are primarily used in katakana. Of the 108 syllables, 37 are generated by simply adding one of the diacritic marks, dakuten (")or handakuten (.sup..smallcircle.), to one of the other 71 syllables. These 71 syllables without diacritic marks can be logically organized into a single matrix of nine or ten rows and eight to ten columns.

Each syllable in the Japanese syllabary consists of either a single vowel, or a consonant followed by a vowel (except only two exceptions). As the number of characters are more than the number of keys in the keyboard the input is done through a multi-stroking system. However, typing standard Japanese text, which includes both kana and kanji (Chinese characters) is a challenging problem.

## 1.2 Method of inputting Japanese characters

Two methods are commonly used to input Japanese to the computer. One, the “kana input method” that uses a keyboard having 46 Japanese iroha (or kana) letter keys. The other method is “Roma-ji input method”, where the Japanese letters are composed of English input from a standard QWERTY keyboard. Both the methods have their advantages and disadvantages.

## 1.3 Criteria for effective inputting of Japanese language

An effective keyboard input system for Japanese language requires the following considerations.

- Arranging the syllables of the Japanese language (kana) on the keyboard.
- The system must minimize the number of keystrokes required to enter text in order to enhance the efficiency of the keyboard.
- The system must reduce the load on the user by reducing the amount of attention and decision making exercise required during the input process.
- The approach should minimize the memory and processing resources required for the purpose.

## 2. Inventions on Japanese keyboards

### 2.1 Keyboard for inputting Japanese characters to computer (Patent 5936556)

**Background problem**

Inputting Japanese characters through a keyboard is a difficult task. Two methods are commonly used to input Japanese to the computer. One, the “kana input method” that uses a keyboard having 46 Japanese iroha (or kana) letter keys. The other method is “Roma-ji input method”, where the Japanese letters are composed of English input from a standard QWERTY keyboard. The first method is not efficient because there are far more kana characters in Japanese. The second method is not efficient as the layout is designed for English input. There is a need to input Japanese character more efficiently.

**Solution provided by the invention**

Sakita invented a novel keyboard inputting system (Patent 5936556, Aug 99) with gojyuon-based key arrangements for inputting Japanese letters. The invention uses a gojyuon table, which shows notations for both the Japanese gojyuon and English alphabet letter. Each letter of the English alphabet is assigned to the key that represents the same (or a similar) sound in the Japanese gojyuon row or column. The software used in conjunction with the keyboard processes the input data, and displays the converted input data on the computer display.

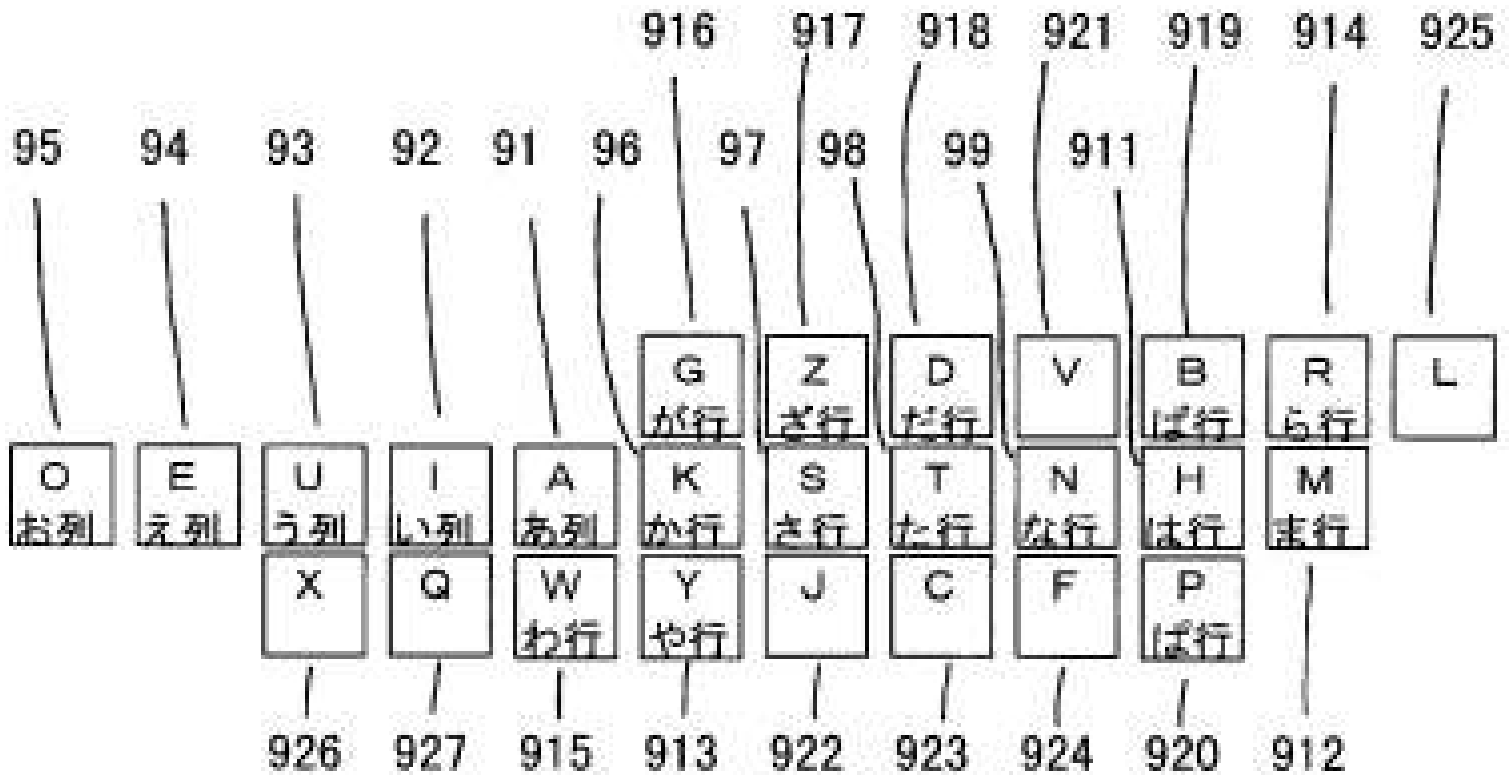

**TRIZ based analysis**

The invention uses a gojyuon table (or conversion table), which stores Japanese gojyuon and corresponding English equivalent, and a software to convert English character combinations to Japanese language **(Principle-24: Intermediary)**.

When the user presses a key on the standard English keyboard it generates a corresponding Japanese character **(Principle-28: Mechanics substitution)**.

## 2.2 Reduced keyboard text input system for the Japanese language (Patent 6646573)

### Background problem

A portable computer does not have enough space for all the keys of a full size keyboard. It is necessary to develop a keyboard that is both small and operable with one hand while capable of entering text into the computer.

The hiragana characters can be entered on a reduced key set keyboard by pressing two or more keystrokes for each kana. But this method is complicated to learn and use.

With the current multi-stroking approach generating a single kana syllable requires an average of at least three keystrokes. Similarly syllables with palatalized vowels can require up to eight keystrokes to generate.

The other alternative approach to enter hiragana on a reduced keyboard is to enter each hiragana with a single keystroke. Each key of the reduced keyboard is associated with multiple hiragana characters. As a user enters a sequence of keys, there is therefore ambiguity in the resulting output since each keystroke may indicate one of several hiragana. There are several approaches by which the system helps the user to resolve the ambiguity of the keystroke sequence.

Thus there is a need to develop a reduced keyboard system that tends to minimize the number of keystrokes required to enter hiragana, and is also simple and intuitive to use.

### Solution provided by the invention

Kushler et al. provided a method of disambiguating to resolve the ambiguities in keystrokes to enter text in Japanese language (Patent 6646573, Assignee: America Online Inc., Issued Nov 2003). The invention uses a word level disambiguation method.

According to the invention, the keyboard may have twelve keys arranged in three columns and four rows as on a standard telephone keypad. A plurality of kana characters and symbols are assigned to each key so that the keystrokes by a user are ambiguous. A user enters a keystroke sequence wherein each keystroke is intended to be the entry of one kana. Each keystroke is intended to represent the phonetic reading of a word or phrase (Yomikata). Because the individual keystrokes are ambiguous, the keystroke sequence could potentially match more than one Yomikata with the same number of kana.
The keystroke sequence is processed by comparing with the keystroke sequence of each Yomikata stored in the dictionary or vocabulary module.

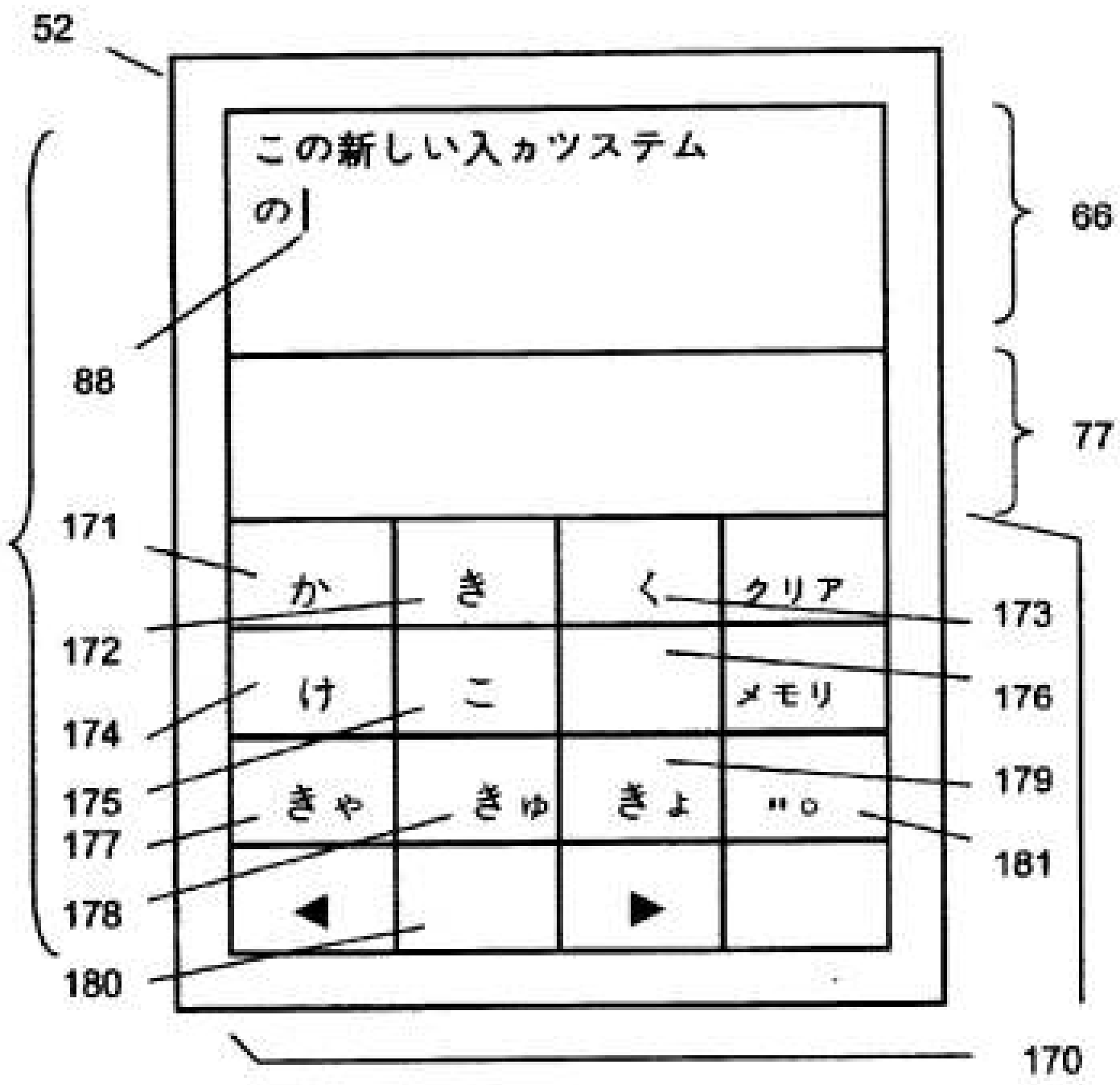

On devices with sufficient display area the selection list may show various Yomikata corresponding to the input sequence to choose from.  But in devices with limited display area, the list of Yomikata is maintained internally and the Yomikatas in the list are displayed one after other when the user presses the same key repeatedly. Finally the user can press the select key to delimit an entered keystroke sequence. After receiving the select key, the disambiguating system automatically selects the most frequently occurring Yomikata in the form of hiragana.

**TRIZ based analysis**

The invention tries to reduce the difficulties of entering Japanese characters from a few numbers of keys by using a word level disambiguation method **(Principle-8: Counterweight)**.

According to invention several (more than one) kana characters are assigned to each individual key on the keyboard **(Principle-17: Another dimension)**.

Ideally the keyboard should predict the Yomikata based on the initial keystroke or kana entered by the user. Because each keystroke could match more than one Yomikata the system filters all Yomikatas to display that matches with the entered keystroke/s **(Principle-16: Partial or excessive action)**.

## 3. Summary and conclusion

There are some interesting peculiarities of Japanese language, which create concerns in inputting and processing Japanese characters and Japanese language into computer.

- Firstly, the Japanese language employs several different alphabets, which may be used in combination.

- Second, Japanese text is typically written without any spaces between words.

- Third, the Japanese language can undergo significant spelling changes to indicate case, tense, politeness, aspect, mode or voice etc.

While inputting Japanese characters through a keyboard remains a challenge, inputting through a reduced sized keyboard like that of a mobile phone becomes even a greater challenge.